\pdfoutput=1
\documentclass[aps,pra,10pt,twocolumn,showpacs,superscriptaddress,nofootinbib]{revtex4-1}
\usepackage{amsmath}
\usepackage{latexsym}
\usepackage{amssymb}
\usepackage{bm}
\usepackage{color}
\usepackage{amsmath}
\usepackage[usenames,dvipsnames,svgnames]{pstricks}
\usepackage{epsfig}
\usepackage{pst-grad} % For gradients
\usepackage{pst-plot} % For axes
\usepackage{newlfont}
\usepackage{amsfonts}
\usepackage{amsthm}
\usepackage{graphicx}
\usepackage[breaklinks]{hyperref}
\usepackage{tikz}

\begin{document}
\title{Bound on efficiency of heat engine from uncertainty relation viewpoint}
\author{Pritam Chattopadhyay}
\email{pritam.cphys@gmail.com}
\affiliation{Cryptology and Security Research Unit, R.C. Bose Center for Cryptology and Security,\\
Indian Statistical Institute, Kolkata 700108, India}

\author{Ayan Mitra}
\email{amitra@lpnhe.in2p3.fr}
\affiliation{Department of Mechanical and Aerospace Engineering, School of Engineering and Digital Sciences, Nazarbayev University,\\ Kabanbay Batyr Ave. 53, Nur-Sultan 010000, Kazakhstan}

\author{Goutam Paul}
\email{goutam.paul@isical.ac.in}
\affiliation{Cryptology and Security Research Unit, R.C. Bose Center for Cryptology and Security,\\
Indian Statistical Institute, Kolkata 700108, India}

\author{Vasilios Zarikas}
\email{vasileios.zarikas@nu.edu.kz}
\affiliation{Department of Mechanical and Aerospace Engineering, School of Engineering and Digital Sciences, Nazarbayev University,\\ Kabanbay Batyr Ave. 53, Nur-Sultan 010000, Kazakhstan}

\pacs{}

% \date{\today}

\begin{abstract}
Quantum cycles in established heat engines can be modeled with various quantum systems as working substances. For example, a heat engine can be modeled with an infinite potential well as the working substance to determine the efficiency and work done. However, in this method, the relationship between the quantum observables and the physically measurable parameters---i.e., the efficiency and work done---is not well understood from the quantum mechanics approach. A detailed analysis is needed to link the thermodynamic variables (on which the efficiency and work done depends) with the uncertainty principle for better understanding. Here, we present the connection of the sum uncertainty relation of position and momentum operators with thermodynamic variables in the quantum heat engine model. We are able to determine the upper and lower bounds on the efficiency of the heat engine through the uncertainty relation.
\end{abstract}

\maketitle 

%%%%%%%%%%%%%%%%%%%%%%%%%%%%%%%%%%%%%%%%%%%%%%%%%%%%%%%%%%%%%%%%%%%%%%%%%%%%%%

\section{Introduction}
Thermodynamics is a prominent theory in evaluating the performance of the engines. It stands as a pillar of theoretical physics and even contributes to our understanding of modern theories, e.g., black hole entropy and temperature~\cite{hawk}, gravity~\cite{thanu,gurs}. Though it is successful in the classical regime, the application of thermodynamics needs to be reinvestigated in a quantum system, as the energy is discrete instead of continuous.  So, we can expect new thermodynamic effects to come up in the quantum world. However, while dealing with thermodynamic laws in a quantum regime, a striking similarity between the quantum--thermodynamic system and the macroscopic system (which are described by classical thermodynamics) can be found. For example, in the case of thermal baths, the Carnot efficiency of the engines is equally relevant for the quantum system (even with a single particle)~\cite{ra}. This raises a question: can all the thermodynamic effects of heat engines of small quantum systems be analyzed and predicted by the known classical thermodynamics? Various works have been performed on the analysis of generic thermodynamic effects and dynamical behavior, which are purely quantum in their nature, with no classical counterpart involved~\cite{uzd}.

Quantum thermodynamics explores thermodynamic quantities like temperature, entropy, heat, etc. for the microscopic system. It can even analyze the above parameters for a single particle model. The study of quantum thermodynamics comprises of the analysis of quantum thermal machines in the microscopic regime~\cite{kos,rezek,rkos1,skr,mkol,htq,jro,lac,rdo} and also in the thermalization mechanism~\cite{arie}. All of the various methods specified so far do not exploit quantum effects in thermodynamics, i.e., there exists some classical analog in these methods. 

A framework for quantum heat engine realization and its  experimental  setup  has  been  proposed~\cite{adec,oab,kzh,ama,wh}. Heat engines can be discrete or continuous in nature. Two-stroke and four-stroke engines belong to the discrete group whereas turbines belong to the continuous engine. 
The Szilard engine was a seminal work~\cite{lsz} to solve the violation of the \mbox{2nd law} of thermodynamics by Maxwell's demon.
The quantum version of this engine was proposed by Kim et al.~\cite{kim}. This is an example of the quantum version of a four-stroke engine. During the insertion and deletion of the barrier in the quantum Szilard engine, a certain amount of work and heat exchange occurs in the system that does not occur in the classical system.  Different models and methods to explain the working principle of the Szilard engine have been explored in various works~\cite{li,kh,cy,lia,jb}. Various models and working mediums for the analysis of thermal machines has been recently explored in various works~\cite{rev1,rev2,rev3,rev4,rev5,rev6,rev6a,rev7,rev8,rev9,rev10,rev11}.

In this work, we study the engines from a more fundamental concept of quantum mechanics and try to connect the efficiency of engines with the fundamental uncertainty relation of two incompatible operators. We consider the one-dimensional potential well as the working substance for the quantum heat engine. Here, we consider a specific model for our analysis, though it is applicable globally to all the engines. We develop an effective method to analyze the useful work using the uncertainty relation of the position and the momentum of a particle in a box without performing any measurement, but by applying two reservoirs of different temperatures. The {}{thorough} analysis performed in this work is done in the nonrelativistic limit. A parallel analysis in the relativistic limit is analyzed in our work \cite{pc23}. 
During the evolution of quantum information, the essence and importance of uncertainty relation in technology got enriched. It has various applications in quantum technology like quantum cryptography~\cite{koas,koas1,caf}, entanglement \mbox{detection~\cite{hofm,ost,mart,guh}}, and even in quantum metrology~\cite{giov} and quantum speed limit~\cite{deba,marv, deff,pires}. In recent times, the \mbox{work~\cite{xiao,maw,baek}} authenticated the uncertainty relation experimentally. The thermal uncertainty relation that we will be applying here is a special case of the general quantum uncertainty relations. The uncertainty relation of two incompatible observables is given by
\begin{equation}
{\Delta}a{\Delta}b{\sim}\frac{\hbar}{2},
\end{equation}  
where $a$ and $b$ are any two canonical variable pairs. No better lower bound was known to us until it was explored in the work~\cite{mondal}. They have not only given a better lower bound than the previously known Pati--Maccone uncertainty relation (PMUR), but also developed an upper bound for the uncertainty relation. It is popularly known as the reverse uncertainty relation.  Using this principle, we will similarly develop the bound of efficiency and work of the heat engine in terms of uncertainty relation.   

We have organized the paper as follows. In Section~\ref{section1}, we develop the thermal uncertainty relation for a one-dimensional potential well of length $2L$. In Section~\ref{section2}, the bound on the sum of variance from the thermal standpoint as well as the traditional one is established. We have devoted Section~\ref{section3} to develop the correlation between the thermodynamic variables and the sum of variance of the position and momentum operator of one-dimensional potential.  Section~\ref{section4} is dedicated to discussing the Stirling cycle and establishing the work done and efficiency in terms of the thermal uncertainty relation. In this section, we illustrate the bound on the work done and efficiency of the quantum engine. The paper is concluded in Section~\ref{section5} with some discussions and future prospects of this work in the field of quantum thermodynamics.

\section{Thermal Uncertainty Relation}\label{section1}
In the first phase of our analysis, we evaluate the thermal uncertainty relation (which is one of the special cases of the general uncertainty relation) for a particle in one-dimensional potential well. To do so, let us first revisit our textbook problem of the one-dimensional potential well.
The one-dimensional potential well is a well-known problem in quantum mechanics~\cite{schiff, dj}.
Here, we consider a particle of mass $m$ inside a one-dimensional potential box of length $2L$. The wave function of this system for the $n$-th level is 
\begin{equation}\label{rev-eq1}
|\psi_n\rangle =\sqrt{\frac{1}{L}} sin(\frac{n\pi x}{2L}).
\end{equation}
So, when the wave function of the model under study is known, we can calculate the eigenvalue of the system. Eigenvalues of the 1-D potential well are
\begin{equation}\label{rev-eq2}
E_n = \frac{n^2\pi^2 \hslash^2}{2m(2L)^2},
\end{equation}
 where $\hslash$ is Planck's constant.

With the wave function of the model in hand, we are all set to derive the uncertainty relation of the position and the momentum for this system. The uncertainty relation for our model is described as~\cite{schiff, dj} 
\begin{eqnarray}\label{c}
\Delta x \Delta p & = & \frac{\hbar}{2} \sqrt{\Big(\frac{(n\pi)^2}{3}-2\Big)}  
\quad \geq   \frac{\hbar}{2},
\end{eqnarray}
where $\Delta x^2 = \langle x^2 \rangle - \langle x \rangle^2$ and $\Delta p^2 = \langle p^2 \rangle - \langle p \rangle^2$ and we have $\langle p \rangle = 0$  for all eigenstates. 
The expectation values of  $\langle x \rangle$, $\langle x^2 \rangle$ and $\langle p^2 \rangle$ can be easily evaluated by considering the defined wave function of the considered system.

We formulate the uncertainty relation of the system at a certain temperature $T$ from a thermodynamics viewpoint. The formulation of the thermal uncertainty relation is performed by the analysis of the partition function of the system. The partition function~\cite{reif1}, $Z$, for the 1-D potential well is expressed as 
\begin{equation}\label{rev-eq3}
 Z = \sum_{n=1}^{\infty} e^{-\beta E_n} \approx \frac{1}{2} \sqrt{\frac{\pi}{\beta \alpha}}\, ,
 \end{equation}
  where $\beta =\frac{1}{k_BT}$, $k_B$ is Boltzmann's constant and $\alpha = \frac{\pi^2 \hbar^2}{2m (2L)^2}$.  The expression of $Z$ converges to the form mentioned, as the product of $\beta$ and $\alpha$ is a small quantity. We use the Gaussian integral as the approximation considering that the error in the integration is negligible as the product of  $\beta$ and $\alpha$ is a small quantity. The mean energy for this system evolves to $\langle E \rangle = - \partial ln Z \big/ \partial \beta = \frac{1}{2 \beta}$. The average of the quantum number for the system under study can be conveyed as
  \begin{equation} \label{rev-eq4}
  \bar{n} = \frac{\sum_n n e^{-\beta E_n}}{\sum_n e^{-\beta E_n}} \approx \frac{1}{\sqrt{\pi \alpha \beta}}\, .
  \end{equation}

Having the mathematical form of the partition function in hand, we have all the resources to develop the thermal uncertainty relation.
Now, we focus on the development of the dispersion relation of the position and the momentum operator at a certain temperature. The dispersion in position can be expressed as
\begin{eqnarray} \label{h} \nonumber
 (\Delta X)^2_T & = & \langle (\Delta X)^2 \rangle_T = \langle X^2\rangle_T - \langle X \rangle^2_T \\ \nonumber 
 & = & \frac{L^2}{3} - \frac{2L^2}{\pi^2} \times \frac{e^{-\alpha\beta} - \sqrt{\pi \alpha \beta} \times  erfc(\sqrt{\alpha\beta})}{\frac{1}{2} \sqrt{\frac{\pi}{\alpha \beta}}} \\  
 & = & \frac{L^2}{3} - \frac{4L^2 \sqrt{\alpha \beta}}{\pi^{5/2}} \times (e^{-\alpha \beta}- \sqrt{\pi \alpha\beta}), 
\end{eqnarray}%MDPI: please check if need indent. NO
$erfc$ is the complementary error function, which appears while solving  $\langle x^2\rangle$.
The dispersion relation of the momentum operator can be analyzed similarly. It is expressed as
\begin{eqnarray} \label{i} \nonumber
 (\Delta P)^2_T & = & \langle (\Delta P)^2 \rangle_T  =  \langle P^2\rangle_T - \langle P \rangle^2_T \\ 
 & = & \frac{\pi^3 \hbar^2 \bar{n}^2}{8 L^2}.  
\end{eqnarray}
 
So, the thermal uncertainty relation  for the system at temperature $T$ can be evaluated from Equations \eqref{h} and \eqref{i} as
\begin{eqnarray} \label{j} \nonumber
 (\Delta X)_T (\Delta P)_T 
 & = & \frac{\hbar \bar{n} \pi^{3/2}}{2\sqrt{2}} {\Big[\frac{1}{3}-\frac{4 }{\bar{n}\pi^{3}}(e^{- \frac{1}{\pi \bar{n}^2}}-\frac{1}{\bar{n}})\Big]}^{\frac{1}{2}} \\ 
 & \geq & \frac{\hbar}{2}. 
\end{eqnarray}

The product uncertainty relation loses its importance when the system under consideration is an eigenstate of the observable under study. The sum of uncertainty~\cite{macc} was introduced to capture uncertainty in the observables when the system happens to be an eigenstate of the observables.
The sum of uncertainty for this system at a particular temperature $T$ is expressed as 
\begin{eqnarray} \label{j1} \nonumber
(\Delta X)_T + (\Delta P)_T 
 & = & L {\Big[\frac{1}{3}-\frac{4 \sqrt{\alpha \beta}}{\pi^{5/2}}(e^{-\alpha \beta}-\sqrt{\pi \alpha \beta})\Big]}^{\frac{1}{2}} \\
 & + &  \frac{\hbar \bar{n} \pi^{3/2}}{2\sqrt{2} L}. 
\end{eqnarray}
The parameters that are considered for the analysis are expressed in table :

\begin{center}
 \begin{tabular}{||c c||} 
 \hline
 Parameter  & Values \\ [0.5ex] 
 \hline\hline
 $\bar{n}$ & 1, 2 \\ 
 \hline
  Length ($L$) & 0-5 nm \\
 \hline
 Hot bath ($T_1$) & 320K \\
 \hline
 Cold bath ($T_2$) & 80K \\ [1ex] 
 \hline
\end{tabular}
\end{center}

\begin{figure}[h]
%\center
  \includegraphics[width=.95\columnwidth]{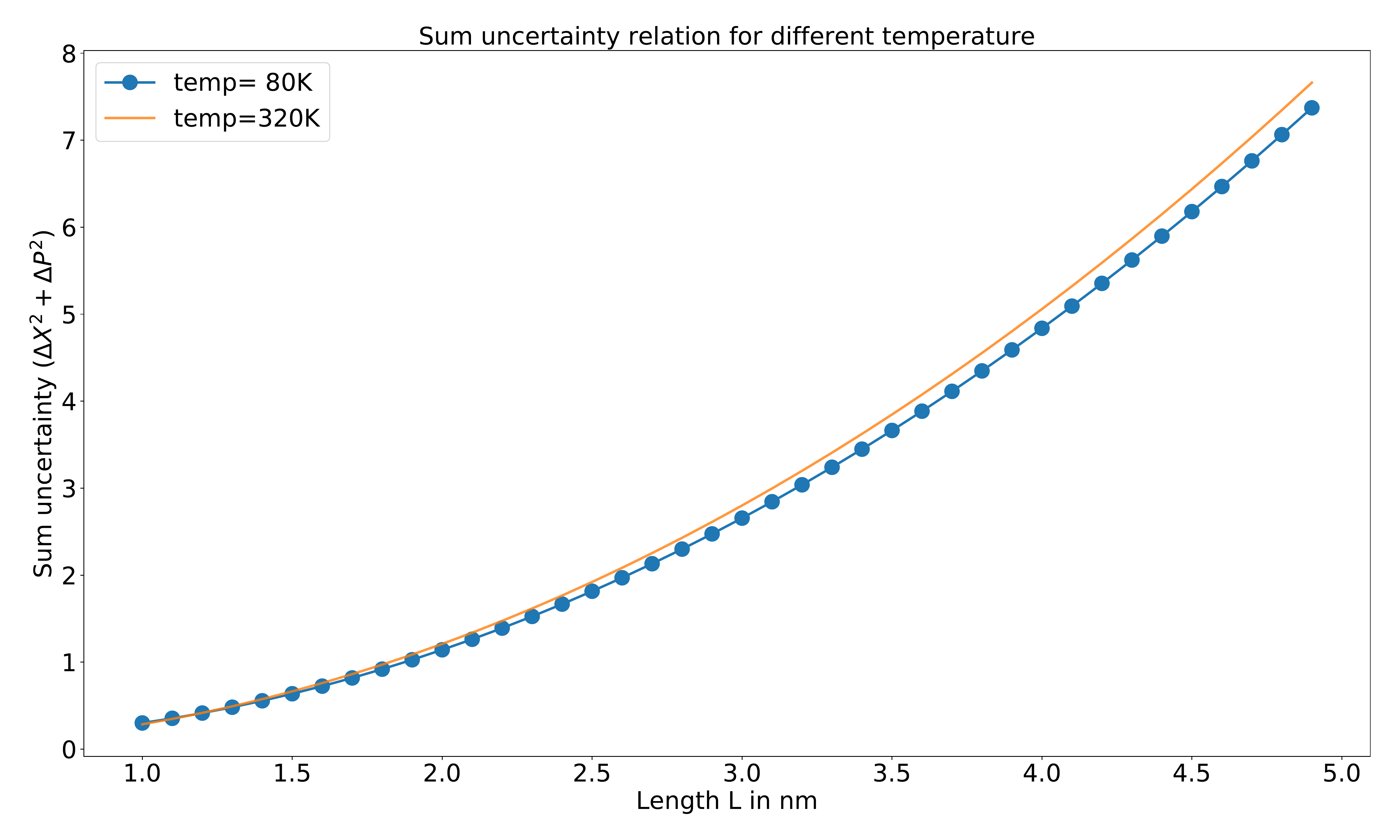} %please correct spacing before and after "=" sign in figure image.
  \caption{The variation of sum uncertainty relation (Equation~\eqref{j1}) for different temperatures. The dotted line is for lower and the solid line is for higher temperature.}
  \label{fig1}
  \end{figure}

  In Figure~\ref{fig1}, the variation of sum uncertainty relation (Equation~\eqref{j1}) with respect to different temperatures is shown. The thermal uncertainty relation that we have developed (Equation~\eqref{j1}) for the considered system encounters a negligible amount of variation when the length of the potential well is small, whereas the difference is large for higher values of $L$ (length %please confirm if intended meaning is retained.  CONFIRMED
  is considered in nanometers).

\vspace{-6pt}

\begin{figure}[h]
%\center
  \includegraphics[width=.95\columnwidth]{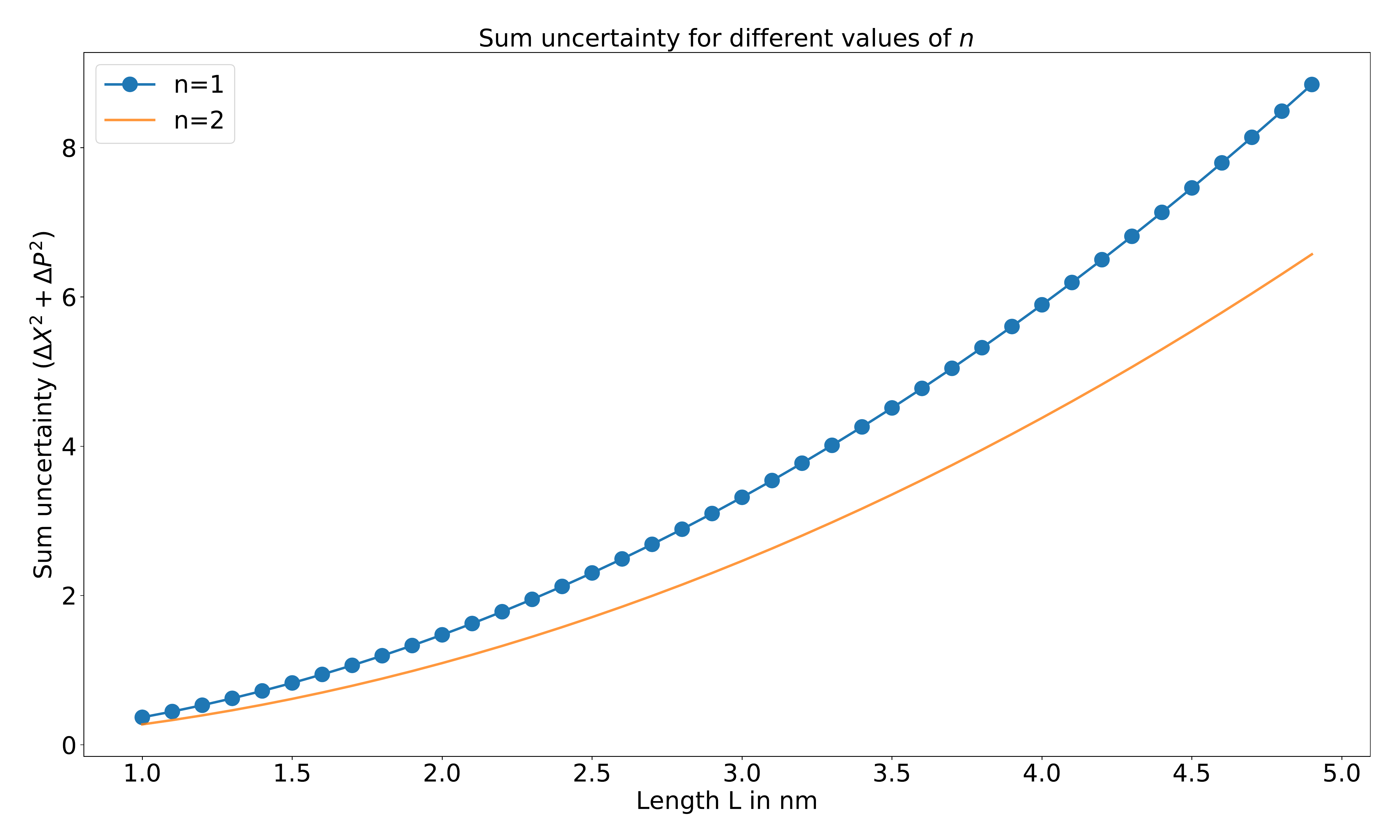}
  \caption{Similarly, this shows variation for different values of $n$.}
  \label{fig2}
  \end{figure}

The variation of uncertainty relation for different levels is shown in Figure~\ref{fig2}. Similar to the case of temperature analysis, the variation is negligible for lower values of $L$ and is large for higher values.
\vspace{-6pt}

\section{Bound on Sum Uncertainty for the One-Dimension Potential Well}\label{section2}

The product of variances is sometimes unable to capture the uncertainty for two incompatible observables. If the state of the system is an eigenstate of one of the observables, then the product of the uncertainties vanishes~\cite{pc,mondal}. To overcome this, the sum of variances is introduced to capture the uncertainty of two incompatible observables. For any quantum model, the sum of variance of two incompatible observables that results in the lower bound  is defined as
\begin{equation} \label{k}
\Delta A^2 +\Delta B^2 \geq \frac{1}{2} \sum_n \Big(\Big|\langle\psi_n|\bar{A}|\psi\rangle\Big|+\Big|\langle\psi_n|\bar{B}|\psi\rangle\Big|\Big)^2.
\end{equation}
For our system, we calculate the lower bound of sum uncertainty for the position and momentum operator. So, we replace $A=X$ and $B=P$, which yields to 
\begin{equation} \label{l}
\Delta X^2+ \Delta P^2 \geq \frac{1}{2} \sum_n \Big(\Big|\langle\psi_n|\bar{X}|\psi\rangle\Big|+\Big|\langle\psi_n|\bar{P}|\psi\rangle\Big|\Big)^2.
\end{equation}
The computation of the reverse uncertainty relation of two observables results to the upper bound of uncertainty relation. To develop the upper bound, we have to utilize the definition of the Dunkl--Williams inequality~\cite{dn1}. The mathematical form of this inequality is 
\begin{equation} \label{m}
\Delta A + \Delta B\leq \frac{\sqrt{2}\Delta (A-B)}{\sqrt{1-\frac{Cov(A,B)}{\Delta A. \Delta B}}}\, .
\end{equation}
 Squaring both sides of the Equation~\eqref{m}, we get the upper bound of the sum of variance for two variables as
\begin{equation}\label{n}
\Delta A^2 + \Delta B^2 \leq \frac{2 \Delta(A-B)^2}{1-\frac{\textrm{Cov}(A,B)}{\Delta A \Delta B}} - 2 \Delta A \Delta B \, ,
\end{equation}
where $\textrm{Cov}(A,B)$ is defined as $Cov(A,B)=\frac{1}{2}\langle \{A,B\} \rangle-\langle A \rangle \langle B \rangle,$ and  $\Delta (A-B)^2= \langle (A-B)^2 \rangle - \langle (A-B) \rangle^2$. $\Delta (A-B)^2$ is the variance of the difference of the two incompatible observables.

For our one-dimensional potential well system, which we have considered as a working substance, we calculate the upper bound of the sum of variance for the position and momentum operator. So, we have to replace $A=X$ and $B=P$ in  Equation~\eqref{n} and it results to 
\begin{eqnarray}\label{q} \nonumber
\Delta X^2 +\Delta P^2 & \leq & \frac{2 \Delta(X-P)^2}{1-\frac{Cov(X,P)}{\Delta X \Delta P}}- 2 \Delta X \Delta P \\ 
& \leq & \frac{L^2}{3} - \frac{2L^2}{(n\pi)^2} + \frac{\pi^2\hbar^2 n^2}{4 L^2}\, .
\end{eqnarray}

In the above equation, i.e., Equation \eqref{q}, the upper bound of the system from the standard  method is developed using the definition described in Equation~\eqref{n}. Now, we develop the sum of variance of the same incompatible observables from the thermodynamic standpoint. The expression for the sum of variance of the system at a particular temperature evolves as
\begin{eqnarray}\label{q1} \nonumber
\Delta X^2 +\Delta P^2  
& \leq & \frac{4L^2}{3} - \frac{8L^2 \sqrt{\alpha \beta}}{\pi^{5/2}} \times (e^{-\alpha \beta}- \sqrt{\pi \alpha\beta}) \\
& + & \frac{\hbar^2\bar{n}^2 \pi^3}{4L^2}\, . 
\end{eqnarray} 

\begin{figure}[h]
%\center
  \includegraphics[width=\columnwidth]{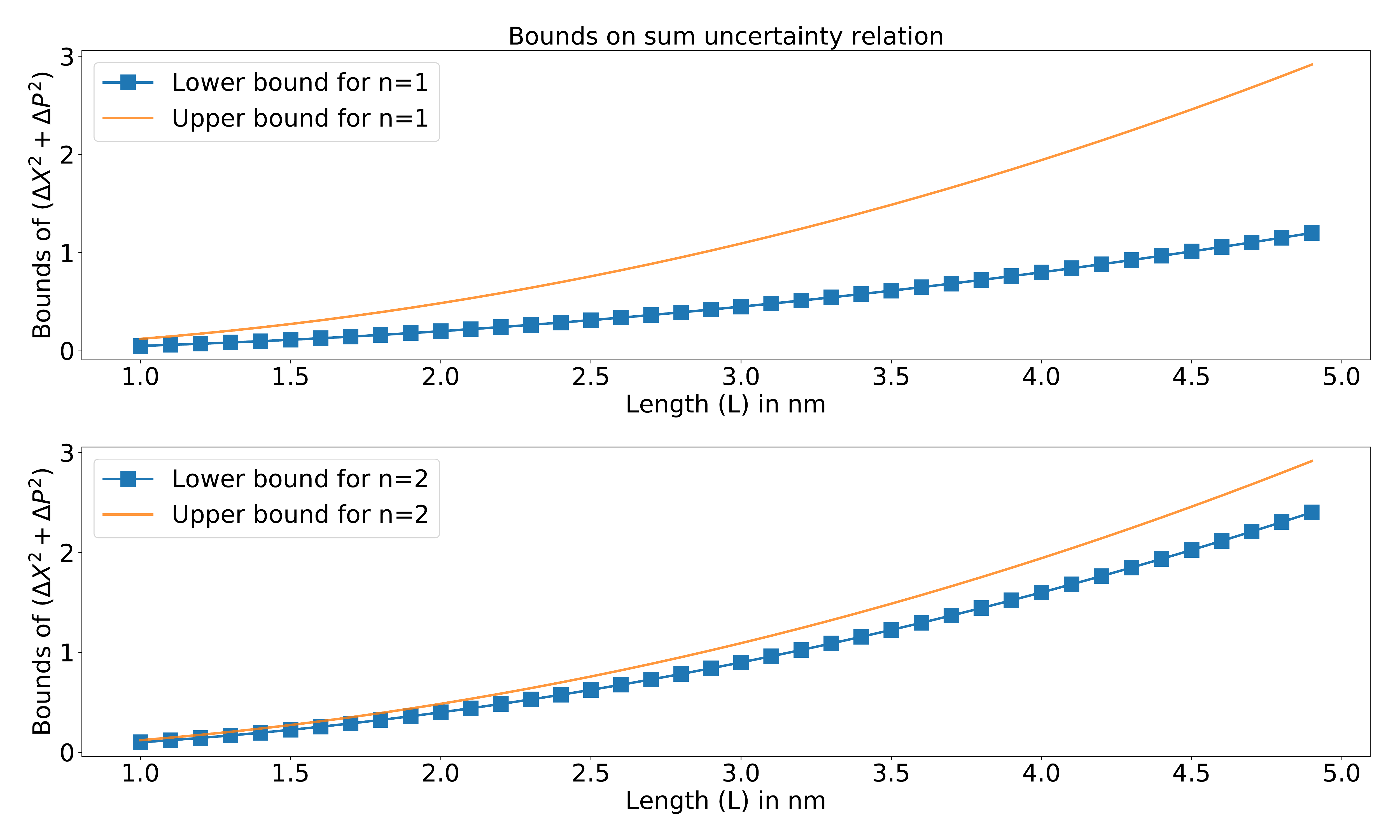}
  \caption{The bound of uncertainty relation (Equation~\eqref{l} and~\eqref{q1}) for a particular temperature for different values of $n$ from a thermodynamic standpoint.}
  \label{fig3}
  \end{figure}

The bounds of sum uncertainty relation (from the thermodynamic perspective developed using Equation~\eqref{l} for the lower bound and Equation~\eqref{q1} for the upper bound for the considered system) with a particular temperature for different levels are %please confirm if plural is correct for "bounds".  CONFIRMED
shown in Figure~\ref{fig3}. The upper part of the plot is for $n=1$ and the lower part is for $n=2$. From Figure~\ref{fig3}, we can infer that the effect of the bounds of uncertainty relation are prominent for higher values of the length of the potential well. The bounds are less prominent for lower values of $L$. 

\vspace{-6pt}

\section{Connection of Thermodynamic Quantities with Uncertainty} \label{section3} 
In the next phase of our analysis, we want to establish a bridge between the thermodynamic quantities with the uncertainty relation. We consider the sum of variance to overcome the flaw that will appear if we consider the product form of uncertainty and if the system is an eigenstate of the observables. We will first demonstrate a connection of partition function with our uncertainty relation. The mathematical form of this is given by
\begin{eqnarray}\label{eq1} 
Z & = & \frac{\pi \bar{n}}{2} 
 =  \frac{L\sqrt{2} }{\hbar \sqrt{\pi}} \Big(\Delta X_T + \Delta P_T + C_T \Big),
\end{eqnarray}
where $C_T = - \frac{L}{3} + \frac{2L}{\pi^{5/2}\sqrt{\alpha \beta_T}} [\alpha \beta_T-\sqrt{\pi}(\alpha \beta_T)^{3/2}-1]$ is a constant for a specific temperature, which is derived by expanding Equation~\ref{j1}, and neglecting the higher order terms as {}{the products of $\alpha$ and $\beta$} are small. 

Since we are to able to bridge a relationship between the uncertainty relation and the partition function, we are set to explain all the thermodynamic variables in terms of uncertainty relations. We develop the Helmholtz free energy~\cite{reif1}, $F$, from the uncertainty viewpoint which takes the form of
\begin{eqnarray} \label{eq3} \nonumber
F & = & \frac{-1}{\beta} ln Z \\ 
& = & \frac{-1}{\beta}\, ln \Big[ \frac{L\sqrt{2} }{\hbar \sqrt{\pi}} \Big(\Delta X_T + \Delta P_T + C_T \Big) \Big]. 
\end{eqnarray}

 Entropy is expressed in terms of Helmholtz free energy. So, we uncover the relationship between the variance of position and momentum with entropy. The mathematical form for entropy from the uncertainty viewpoint is
 \begingroup\makeatletter\def\f@size{10}\check@mathfonts
\def\maketag@@@#1{\hbox{\m@th\normalsize\normalfont#1}}%
\begin{eqnarray} \label{eq4} \nonumber
S & = & - \frac{\partial F}{\partial T} \\ \nonumber
& = & k_B\,\, ln\Big[\frac{L\sqrt{2} }{\hbar \sqrt{\pi}} \Big(\Delta X_T + \Delta P_T + C_T \Big) \Big] \\ \nonumber
 & + &  \frac{\hbar\sqrt{\pi}k_B(\nu +\gamma)}{\sqrt{2} L \beta  (\Delta X_T + \Delta P_T + C_T)}\, ,\\
\end{eqnarray}
\endgroup
where $\nu = \frac{ \frac{L^2 \sqrt{\alpha}}{\sqrt{\beta}\pi^{5/2}} \Big(e^{-\alpha \beta}-\sqrt{\pi \alpha \beta}\Big) - \frac{2L^2\sqrt{\alpha \beta}}{\pi^{5/2}} \Big(\alpha e^{-\alpha \beta}-1/2 \sqrt{\frac{\pi \alpha}{\beta}}\Big)} {\Big[\frac{L^2}{3}-\frac{4L^2\sqrt{\alpha \beta}}{\pi^{5/2}} \Big(e^{-\alpha \beta}-\sqrt{\pi \alpha \beta}\Big) \Big]^\frac{1}{2}}$ and $\gamma$ is  expressed as \linebreak $\gamma = -\frac{L}{\pi^{5/2}\sqrt{\alpha}\beta^{3/2}} (\alpha \beta -\sqrt{\pi}(\alpha \beta)^{3/2}-1) + \frac{2L}{\pi^{5/2}\sqrt{\alpha \beta}}(\alpha- \sqrt{\pi \beta} \alpha^{3/2})$.

%\begin{figure}[H]
%  \includegraphics[scale=0.17]{Entropy.pdf}
%  \caption{The variation of entropy for different values of $n$.}
%  \label{fig4}
%  \end{figure}

\begin{figure}[h]
 % \center
  \includegraphics[width=\columnwidth]{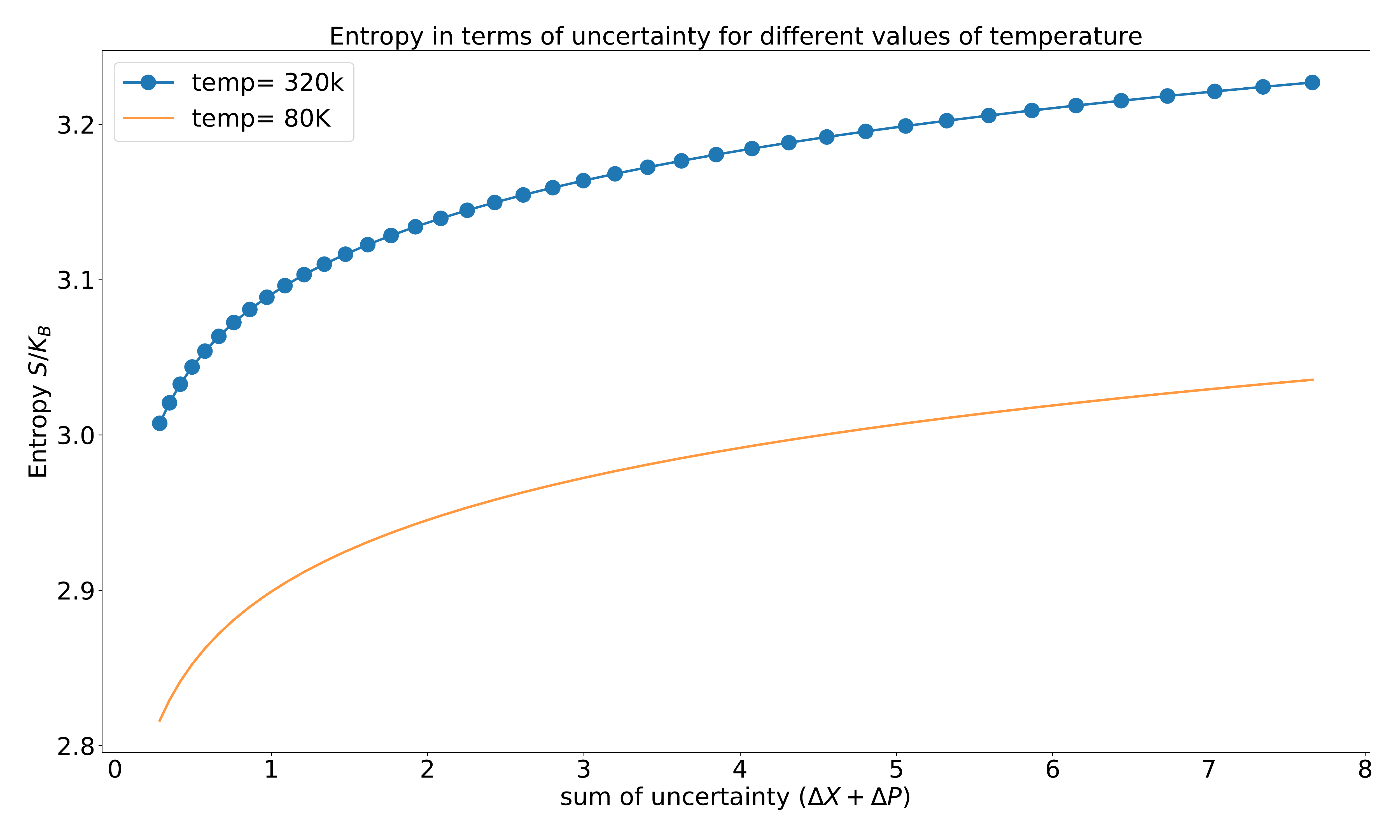}
  \caption{The variation of entropy (Equation~\eqref{eq4}) for different values of temperature. The scattered plot is for higher temperature and solid line is for lower temperature.}
  \label{fig5}
  \end{figure}

In  Figure~\ref{fig5}, the variation of entropy in terms of the uncertainty relation is shown. We can observe an increase in the entropy with an increase in the uncertainty for different temperatures. With an increase in uncertainty, the disorder in the system increases, causing an increase in the entropy.

We know that entropy is a measure of entanglement. So, from this relation, we are able to bridge a connection between the uncertainty relation with entanglement. So, we can measure the entanglement property of the system from the uncertainty relation if we are able to model our system with any well-known quantum systems.

 Knowing the Helmholtz free energy~\cite{saham,reif1} $F$, for a given system, all the relevant thermodynamic quantities can be computed from it. Here, the correlation of $F$ with the uncertainty in the measure of the position and momentum is established. Hence, this replaces the explicit requirement of computing the internal energy of the system for deriving the thermodynamic quantities. In addition, it also raises the question of whether phase transition (Landau theory and its multimode coupling) can be analyzed from an uncertainty perspective.
\vspace{-6pt}

\section{Stirling Cycle and Bound on Efficiency}\label{section4}
\begin{figure}[h]
%\center
 \includegraphics[width=\columnwidth]{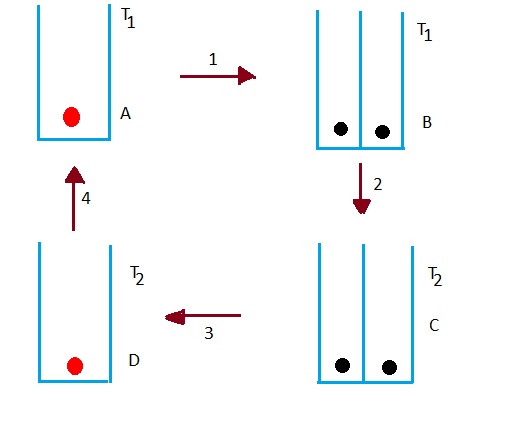}
 \caption{ The figure shows the four stages (two isothermal and two isochoric processes) of the Stirling cycle modeled using the potential well.}
 \label{fig6}
\end{figure}

A Stirling cycle~\cite{say,agar,hua,blick} is composed of four stages, two isothermal processes, and two isochoric processes.
During the first stage, we insert a barrier isothermally in the middle of the well.  While this quasi-static insertion process is being done, the working medium stays at an equilibrium condition with a hot bath at a temperature $T_1$.
During the second stage, we perceive an isochoric heat extraction of the working medium by connecting it with a bath at a lower temperature of $T_2$. In the next stage of the cycle, there is an isothermal removal of the barrier where we retain the engine in equilibrium at temperature  $T_2$. In the final stage, we bridge the engine to the hot bath at temperature $T_1$ and this give rise to isochoric heat absorption. It is represented pictorially in Figure~\ref{fig6}.

Similar to~\cite{thomas}, we calculate the work done and the efficiency but in terms of uncertainty relation. To determine the work done of the engine, a one-dimensional well of length $2L$ is considered with a particle of mass $m$ at a temperature $T_1$. The energy of this system is $E_n = \frac{n^2\pi^2 \hslash^2}{2m(2L)^2}$. The partition function $Z_A$ for the system is $Z \approx \frac{1}{2} \sqrt{\frac{\pi}{\beta \alpha}}$. Then, a wall is inserted isothermally that converts the potential well into an infinite double well potential. Due to this insertion of the wall, the energy level for even values of $n$ remain unchanged but the odd ones shift and overlap with their nearest neighboring energy level. So, the energy of the newly formed partitioned one-dimensional potential box is
\begin{equation}\label{r}
E_{2n} = \frac{(2n)^2\pi^2 \hslash^2}{2m(2{}{L})^2}.
\end{equation}

So, the new partition function stands as
\begin{eqnarray}\label{s}
Z_B = \sum_n 2e^{-\beta_1 E_{2n}}. 
\end{eqnarray}

The internal energies for the system are calculated from the partition function. The internal energy $U_A$ and $U_B$ is defined as $U_i = - \partial ln Z_i \big/ \partial \beta_1$, where $i=A,B$ and $\beta_1= \frac{1}{k_BT_1}$. This results to
\begin{equation}\label{t}
U_A= U_B= \frac{1}{2\beta_1}.
\end{equation}
The heat exchanged in this isothermal process can be expressed as 
\begin{equation}\label{u}
Q_{AB} = U_B - U_A + k_B T_1 ln Z_B - k_B T_1 ln Z_A.
\end{equation}
Then, the system is connected to a heat bath at a lower temperature $T_2$. The partition function for this lower temperature, where the energy remains the same, is defined as
\begin{equation}\label{v}
Z_C = \sum_n 2e^{-\beta_2 E_{2n}}.
\end{equation}
The heat exchanged for this stage of the cycle is the difference  of  the  average  energies  of  the  initial  and  the  final states, i.e.,
\begin{equation}\label{w}
Q_{CB} = U_C- U_B,
\end{equation}
where $U_C= - \partial ln Z_C\big/ \partial \beta_2$ and $\beta_2= \frac{1}{k_BT_2}$. While the system is connected  to  the  heat  bath  at  temperature $T_2$, we remove the wall isothermally, which we call stage 3. The energy is now of the form $E_n = \frac{n^2\pi^2 \hslash^2}{2m(2L)^2}$. The corresponding partition function is given by
\begin{equation}\label{x}
Z_D = \sum_n e^{-\beta E_n}.
\end{equation}
We can calculate the internal energy $U_D$ similarly as $U_C$. The heat exchanged during this process is given by
\begin{equation}\label{y}
Q_{CD} = U_D - U_C + k_B T_2 ln Z_D - k_B T_2 ln Z_C.
\end{equation}
In the fourth stage of the cycle, the system is connected back to the heat bath at temperature $T_1$. The corresponding energy exchange for this stage can be expressed as 
\begin{equation}\label{z}
Q_{DA} = U_A- U_D.
\end{equation}
So, the total work done for the process in terms of variance of the position and the momentum operator evolves to 
\begin{eqnarray}\label{aa} \nonumber
W & = & Q_{AB} + Q_{BC} + Q_{CD} + Q_{DA} \\ 
& = & \frac{8L^2\alpha}{\hbar^2 \pi^2} \Big[D\, ln\Big(\frac{Z_B}{Z_A}\Big)+ E\, ln\Big(\frac{Z_D}{Z_C}\Big)\Big] . 
\end{eqnarray}

The efficiency of the Stirling cycle engine stands as 
\begin{eqnarray} \label{bb}\nonumber
\eta & = & 1 + \frac{Q_{BC} + Q_{CD}}{Q_{DA} + Q_{AB}} \\ \nonumber
& = & \frac{ \Big(\bar{n}_{T_2}^2 \, ln\Big(\frac{Z_D}{Z_C}\Big) + \bar{n}_{T_1}^2 \, ln\Big(\frac{Z_B}{Z_A}\Big)\Big) }{\Big(-\bar{n}_{T_2}^2/2 + \bar{n}_{T_1}^2\Big(ln\Big(\frac{Z_B}{Z_A}\Big)+1/2\Big)\Big)} \\ 
& = & \frac{\Big[D \, ln\Big(\frac{Z_B}{Z_A}\Big) + E \, ln\Big(\frac{Z_D}{Z_C}\Big) \Big]}{\Big[-E/2 + D \Big(\, ln\Big(\frac{Z_B}{Z_A}\Big)+1/2\Big) \Big]}\, , 
\end{eqnarray}
where $D=\frac{8L^2}{\pi^3 \hbar^2}(\Delta X_{T_1} + \Delta P_{T_1} + C_{T_1})^2$ and $E= \frac{8L^2}{\pi^3 \hbar^2} (\Delta X_{T_2} +\Delta P_{T_2} + C_{T_2})^2$. 

In Equation \eqref{bb}, the upper and lower bound of the efficiency are evaluated in terms of the bound that is being analyzed for the thermal uncertainty relation of the position and the momentum operator. Here, the expression of $D$ and $E$ (for the working model considered for the analysis of the heat engine) gives the required uncertainty relation for the illustration of the bound of the efficiency.

\begin{figure}[h]
%\center
  \includegraphics[width=\columnwidth]{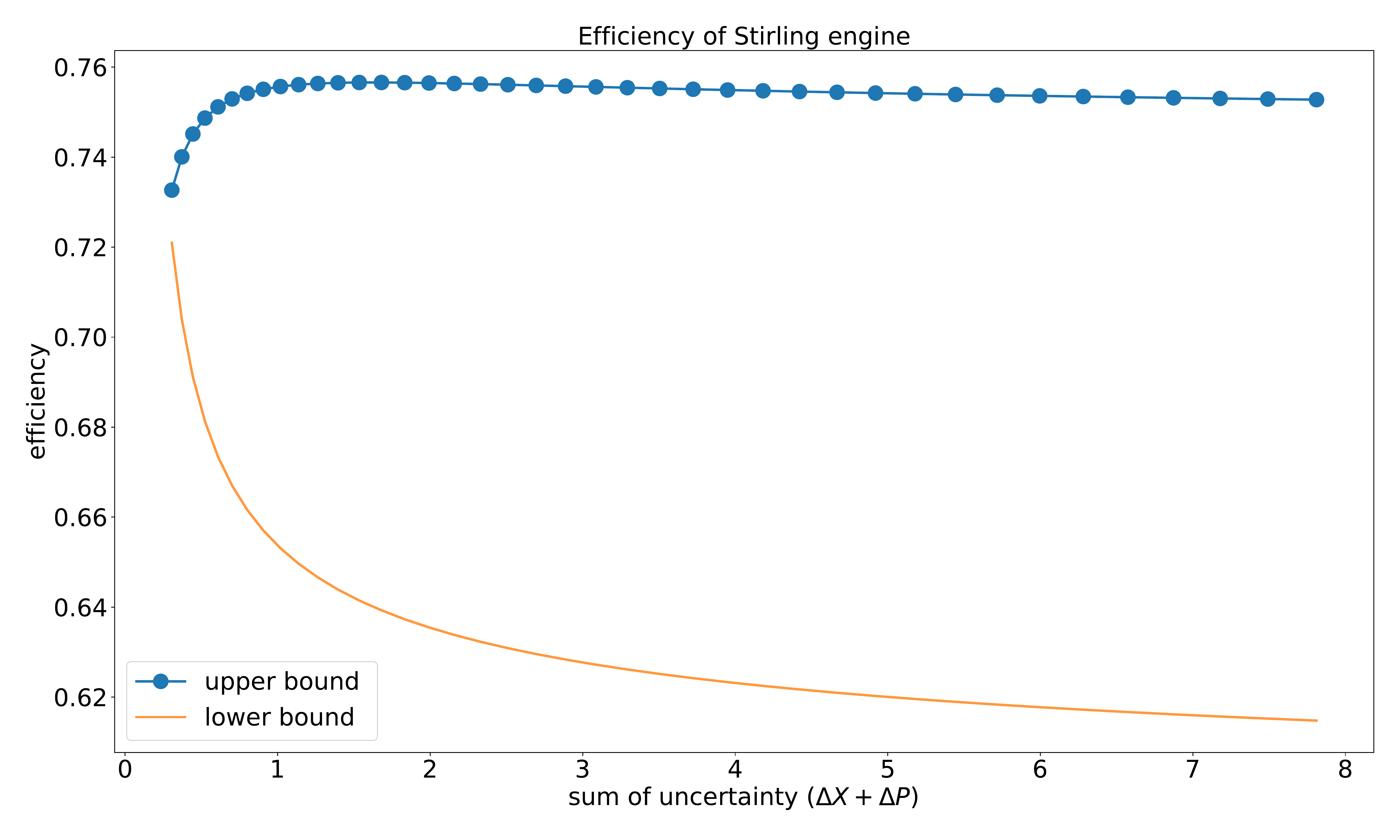}
  \caption{The bounds on the efficiency by heat engine in terms of the uncertainty relation. The dotted line %please confirm if intended meaning is retained. 
  	represents the upper bound and the solid line represents the lower bound of the efficiency.}
  \label{fig8}
  \end{figure}

In this paper, we are able to bridge a connection between the efficiency of the heat engine with the variance of the position and the momentum operator. The upper bound of the efficiency for the heat engine is near-constant when the uncertainty is high, whereas it dips a little when uncertainty is low. As shown in Figure~\ref{fig8}, the lower bound of the efficiency is high when the uncertainty in measurement is less and dips gradually with the increase in uncertainty. Thus, with an increase in uncertainty, we can visualize that the lower bound of the efficiency decreases. From Figure~\ref{fig8}, one can infer that the lower and the upper bound of the efficiency are near about the same when the uncertainty in the position and the momentum operator is quite small. {}{The decrease in the efficiency with the uncertainty is due to the fact that the conversion ratio of the thermal machine decrease with the increase in the thermal  energy of the system, which has a relation with the uncertainty of the working medium. The asymptotic behavior for the higher values of the uncertainty is due to the fact that the conversion rate gets saturated. Computing the error margins would help us  understand this behavior better, which we intend to do in a future paper.}

With the increase in uncertainty, the conversion ratio of the heat engine decreases as the thermal energy of
the system is directly proportional to the uncertainty of the system. In the case of the upper bound of the efficiency,
which is depicted in terms of the uncertainty relation defined in Equation 22, the decrease in the efficiency is more
prominent due to the presence of the exponential component, which causes exponential growth in the thermal energy
of the engine and the dissipated heat over the work output.
 
 The Carnot efficiency for the low temperature limit is expressed as $\Big(1-\frac{T_2}{T_1}\Big)$, where $T_2$ and $T_1$ are the temperature of the cold and hot bath, respectively. The upper bound of the efficiency from an uncertainty viewpoint is consistent with the bound given by Carnot cycle. So, we can infer that the position and the momentum of the particle has a direct link with the thermodynamic variables. The work~\cite{nied} suggests that the efficiency of engines that are powered by nonthermal baths can be higher than the usual convention. This can be testified from an uncertainty viewpoint.
 
In the quantum regime, after measurement, the system collapses to one of its eigenstates. So, to describe and analyze the efficiency of the engine for different conditions, we must have multiple copies of %please confirm if intended meaning is retained.  CONFIRMED
the system, whereas if we know the uncertainty relation of the model considered for analysis, we can describe and analyze different conditions without any measurement being done on the system. So, multiple copies of the system are not required for our analysis. This also reduces the measurement cost for analysis of the system.
\vspace{-6pt}

\section{Discussion}\label{section5}
The quantum heat engine has a predominant role in better understanding of the quantum engines, information, and quantum thermodynamics. This work develops a relationship between the thermodynamic variables with the position and momentum of the particle in the system. We give the analytic formulation of the work and efficiency of the engine in terms of the thermal uncertainty relation. Though we have considered a specific model for our analysis, this analysis can cater to a spectrum of global effects, i.e., it can be used to explain the efficiency of the various engines with different quantum models as the working substance. Based on these formulations, the physical properties of the heat engine and the thermodynamic variables that we have encountered are as follows:

\begin{enumerate}
\item[(a)] Every quantum thermodynamic variable has a direct connection with the uncertainty relation. Helmholtz free energy shows the dependence of the internal energy of the thermodynamic system with the uncertainty relation of the incompatible observables. The detailed analysis of entropy with the uncertainty relation shows that entropy increases when the uncertainty of any one of the observables increases for a definite temperature. The rate of increase is different for different temperatures {}{(Figure \ref{fig5})}.

\item[(b)] The total work and the efficiency depends on the position and momentum of the particle. The change in the uncertainty of the position and the momentum has a direct impact on the efficiency rate and the work of the engine. The lower bound of the efficiency of the engine drops gradually when the uncertainty of the observable increases {}{(Figure \ref{fig8})}. The upper bound of the efficiency {}{(Figure \ref{fig8})} shows a small variation for higher uncertainty relation, which conveys that the conversion rate of work input to output is near-constant for higher uncertainty.

\item[(c)] The uncertainty relation, which is the fundamental principle of quantum mechanics, is able to predict the efficiency and the total work of the engine without performing any measurement. So, the measurement cost for the system gets reduced if we are able to model the system under study with a quantum model for which we can develop the uncertainty relation. 
\end{enumerate}

\section{Conclusions}

The bridge of the uncertainty relation with  thermodynamic variables raises a question of whether we can {}{analyze} the phase transition (Landau theory) from an uncertainty perspective.

 Most of the known methods for the measurement of entanglement converge to the analysis of entropy~\cite{plen}. If we can model the system that is being analyzed with a quantum model, we can construct the entanglement from the uncertainty relation of the system. This would be a standard method to measure the entanglement property of the system, which might be a solution to the open problem of entanglement measure.

The 1-D problem in the nonrelativistic case is a standard problem. However, in the relativistic case it is not a standard problem. The study of the heat engine with a relativistic particle can be analyzed. The mapping of the entropy with uncertainty to explain the entanglement property for the relativistic system~\cite{albe} is an open area to explore. Even the holographic interpretation of entanglement entropy of anti-de Sitter (ADS)/conformal field theory (CFT)~\cite{ryu} can be mapped with uncertainty relation.

This work can be extended to the development of quantum engines in deformed space structures~\cite{kempf,ques,melj} through the correlation of the generalized uncertainty relation and thermodynamic variables. In the paper~\cite{rezek}, the authors mentioned the noncommutativity of kinetic and potential energy of the quantum harmonic heat engine. Therefore, the possibility of a connection between the deformed space structures~\cite{pc} and the heat engines can be further explored in the future. One can even study the anharmonic oscillator models through the uncertainty standpoint.

The study of other thermodynamic cycles and procuring the bound for different thermodynamic parameters is a wide open area to explore. As entropy can be mapped with the uncertainty relation, this leaves us with questions for future study as to whether all thermodynamic analyses can be mapped with the uncertainty of the observables.

\vspace{6pt}

%

%%%%%%%%%%%%%%%%%%%%%%%%%%%%%%%%%%%%%%%%%
%
%%%%%%%%%%%%%%%%%%%%%%%%%%%%%%%%%%%%%%%%%

%%%%%%%%%%%%%%%%%%%%%%%%%%%%%%%%%%%%%%%%%%%%%%%%%%%%%%%%%%%%%%%%%


\begin{thebibliography}{999}
\bibitem{hawk} Hawking, S.W. Black holes and thermodynamics. \emph{Phys. Rev. D} \textbf{1976}, \emph{13}, 191.

\bibitem{thanu} Padmanabhan, T. Gravity and the thermodynamics of horizons. \emph{Phys. Rep.} \textbf{2005}, \emph{406}, 49--125.

\bibitem{gurs} G\"ursoy,U.;  Kiritsis, E.; Mazzanti,  L.; Nitti, F. Holography and thermodynamics of 5D dilaton-gravity. \emph{J. High Energy Phys.} \textbf{2009}, \emph{2009}, 033.

\bibitem{ra} Alicki, R. The Quantum Open System as a Model of the Heat Engine. \emph{J. Phys. A} \textbf{1979}, \emph{12}, L103.

\bibitem{uzd} Uzdin, R.;  Levy, A.; Kosloff, R. Equivalence of quantum heat machines, and quantum-thermodynamic signatures. \emph{Phys. Rev. X} \textbf{2015}, \emph{5}, 031044.

\bibitem{kos} Kosloff, R. A Quantum Mechanical Open System as a Model of a Heat Engine. \emph{J. Chem. Phys.} \textbf{1984}, \emph{80}, 1625.

%\bibitem{rez} Rezek, Y.; Kosloff, R. \hl{Irreversible Performance of a Quantum Harmonic Heat Engine.} \emph{New J. Phys.} \textbf{2006}, \emph{8}, 83. % Removed

\bibitem{rkos1} Kosloff, R.;  Levy, A. Quantum Heat Engines and Refrigerators: Continuous Devices. \emph{Annu. Rev. Phys. Chem.} \textbf{2014}, \emph{65}, 365.

\bibitem{skr} Skrzypczyk, P.;  Short, A.J.; Popescu, S. Work Extraction and Thermodynamics for Individual Quantum Systems. \emph{Nat. Commun.} 2014, 5, 4185.

\bibitem{mkol} Kol\'a\v{r}, M.; Gelbwaser-Klimovsky,  D.;  Alicki, R.; Kurizki, G. Quantum Bath Refrigeration Towards Absolute Zero: Challenging the Unattainability Principle. \emph{Phys. Rev. Lett.} \textbf{2012}, \emph{109}, 090601.

\bibitem{htq}   Quan, H.T.;  Liu, Y.X.;  Sun, C.P.; Nori, F. Quantum Thermodynamic Cycles and Quantum Heat Engines. \emph{Phys. Rev. E} \textbf{2007}, \emph{76}, 031105.

\bibitem{jro} Ro{\ss}nagel, J.;  Abah, O.;  Schmidt-Kaler, F.;  Singer, K.; Lutz,  E. Nanoscale Heat Engine Beyond the Carnot Limit. \emph{Phys. Rev. Lett.} \textbf{2014}, \emph{112}, 030602.

\bibitem{lac} Correa, L.A.;  Palao, J.P.;  Alonso, D.; Adesso, G. Quantum-Enhanced Absorption Refrigerators. \emph{Sci. Rep.} \textbf{2014}, \emph{4}, 3949.

\bibitem{rdo} Dorner, R.;  Clark, S.R.;  Heaney, L.;  Fazio, R.;  Goold, J.; Vedral, V. Extracting Quantum Work Statistics and Fluctuation Theorems by Single-Qubit Interferometry. \emph{Phys. Rev. Lett.} \textbf{2013}, \emph{110}, 230601.

\bibitem{arie} Riera, A.;  Gogolin, C.; Eisert, J. Thermalization in Nature and on a Quantum Computer. \emph{Phys. Rev. Lett.} \textbf{2012}, \emph{108}, 080402.

\bibitem{oab} Abah, O.;  Ro{\ss}nagel, J.;  Jacob, G.;  Deffner, S.;  Schmidt Kaler, F.;  Singer, K.; Lutz, E. Single-Ion Heat Engine at Maximum Power. \emph{Phys. Rev. Lett.} \textbf{2012}, \emph{109}, 203006.

\bibitem{adec} Dechant, A.;  Kiesel, N.; Lutz, E. All-Optical Nano-mechanical Heat Engine. \emph{Phys. Rev. Lett.} \textbf{2015}, \emph{114}, 183602.

\bibitem{kzh} Zhang, K.;  Bariani, F.; Meystre, P. Quantum Opto-mechanical Heat Engine. \emph{Phys. Rev. Lett.} \textbf{2014}, \emph{112}, 150602.
%19
\bibitem{ama} Mari, A.; Eisert, J. Cooling by Heating: Very Hot Thermal Light Can Significantly Cool Quantum Systems. \emph{Phys. Rev. Lett.} \textbf{2012}, \emph{108}, 120602.
%37
\bibitem{wh}  Zurek, W.H. Maxwell’s demon, Szilard’s engine and quantum measurements. \emph{arXiv} \textbf{2003}, arXiv:quant-ph/0301076.
%32
\bibitem{lsz}  Szilard, L. \"Uber die Entropieverminderung in einem thermodynamischen System bei Eingriffen intelligenter Wesen. \emph{Zeitschrift für Physik} \textbf{1929}, \emph{53}, 840--856.
%33
\bibitem{kim}  Kim, S.W.;  Sagawa, T.;  De Liberato, S.; Ueda, M. Quantum Szilard engine.  \emph{Phys. Rev. Lett.} \textbf{2011}, \emph{106}, 070401.
%38
\bibitem{li} Li,  H.;  Zou, J.;  Li, J.-G.;  Shao, B.; Wu, L.-A.  Revisiting the quantum Szilard engine with fully quantum considerations. \emph{Ann. Phys.} \textbf{2012}, \emph{327}, 2955.

\bibitem{kh}  Kim, K.H.; Kim, S.W.  Szilard's information heat engines in the deep quantum regime. \emph{J. Korean Phys. Soc.} \textbf{2012}, \emph{61}, 1187.

\bibitem{cy}  Cai, C.Y.;  Dong, H.; Sun, C.P.  Multiparticle quantum Szilard engine with optimal cycles assisted by a Maxwell's demon. \emph{Phys. Rev. E} \textbf{2012}, \emph{85}, 031114.

\bibitem{lia}  Zhuang, Z.; Liang,  S.D. Quantum Szilard engines with arbitrary spin. \emph{Phys. Rev. E} \textbf{2014}, \emph{90}, 052117.
%42
\bibitem{jb}  Bengtsson, J.;  Nilsson Tengstrand, M.;  Wacker, A.; Samuelsson, P.;  Ueda, M.; Linke,  H.; Reimann, S.M. Supremacy of the quantum many-body Szilard engine with attractive bosons. \emph{arXiv} \textbf{2017}, arXiv:1701.08138.

%20
\bibitem{rev1}Kosloff, R.; Rezek, Y.  The quantum harmonic Otto cycle. \emph{Entropy} \textbf{2017}, \emph{19}, 136.

\bibitem{rev2}Deffner, S.  Efficiency of harmonic quantum Otto engines at maximal power. \emph{Entropy} \textbf{2018}, \emph{20}, 875.

\bibitem{rev3}Bustos-Marún, R.A.; Calvo, H.L.  Thermodynamics and steady state of quantum motors and pumps far from equilibrium. \emph{Entropy} \textbf{2019}, \emph{21}, 824.

\bibitem{rev4}Meng, Z.; Chen, L.; Wu, F.  Optimal power and efficiency of multi-stage endoreversible quantum Carnot heat engine with harmonic oscillators at the classical limit. \emph{Entropy} \textbf{2020}, \emph{22}, 457.

\bibitem{rev5} Insinga, A.R.  The quantum friction and optimal finite-time performance of the quantum Otto cycle. \emph{Entropy} \textbf{2020}, \emph{22}, 1060.


\bibitem{rev6}Dann, R.; Kosloff, R.; Salamon, P.  Quantum finite-time thermodynamics: Insight from a single qubit engine. \emph{Entropy} \textbf{2020}, \emph{22}, 1255.

\bibitem{rev6a}Kosloff, R. Quantum thermodynamics and open-systems modeling. \emph{ J. Chem. Phys.} \textbf{2019}, \emph{150}, 204105.

\bibitem{rev7} Chen, L.; Liu, X.; Wu, F.; Xia, S.; Feng, H.  Exergy-based ecological optimization of an irreversible quantum Carnot heat pump with harmonic oscillators. \emph{Phys. A Stat. Mech. Appl.} \textbf{2020}, \emph{537}, 122597.

\bibitem{rev8} Diaz, G.A.; Forero, J.D.; Garcia, J.; Rincon, A.; Fontalvo, A.; Bula, A.; Padilla, R.V.  Maximum power from fluid flow by applying the first and second laws of thermodynamics. \emph{J. Energy Resour. Technol.} \textbf{2017}, \emph{139}, 3.


\bibitem{rev9}Del Campo, A.; Boshier, M.G.  Shortcuts to adiabaticity in a time-dependent box. \emph{Sci. Rep.} \textbf{2012}, \emph{2}, 1--6.


\bibitem{rev10} Stefanatos, D.  Optimal shortcuts to adiabaticity for a quantum piston. \emph{Automatica} \textbf{2013}, \emph{49}, 3079--3083.
%%39
\bibitem{rev11}Stefanatos, D.  Exponential bound in the quest for absolute zero. \emph{Phys. Rev. E} \textbf{2017}, \emph{96}, 042103.
%%%40
\bibitem{pc23}  Chattopadhyay, P.; Paul, G. Relativistic quantum heat engine from uncertainty relation standpoint. \emph{Sci. Rep.} \textbf{2019}, \emph{9}, 16967.
%63
\bibitem{caf}  Fuchs, C.A.; Peres, A. Quantum-state disturbance versus information gain: Uncertainty relations for quantum information. \emph{Phys. Rev. A} \textbf{1996}, \emph{53}, 2038.


\bibitem{koas} Koashi, M. Unconditional security of quantum key distribution and the uncertainty principle. \emph{J. Phys. Conf. Ser.} \textbf{2006}, \emph{36}, 16.

\bibitem{koas1} Koashi,  M. Simple security proof of quantum key distribution via uncertainty principle. \emph{arXiv} \textbf{2005}, arXiv:0505108.
%%53
\bibitem{hofm}  Hofmann, H.F.; Takeuchi,  S. Violation of local uncertainty relations as a signature of entanglement. \emph{Phys. Rev. A} \textbf{2003}, \emph{68}, 032103.

\bibitem{ost} Osterloh, A. Entanglement and its facets in condensed matter systems. \emph{arXiv} \textbf{2008}, arXiv:0810.1240.

\bibitem{mart}  Marty, O.; Epping, M.; Kampermann,  H.;  Bru{\ss}, D.; Plenio, M.B.; Cramer, M.  Quantifying entanglement with scattering experiments. \emph{Phys. Rev. B} \textbf{2014}, \emph{89}, 125117.
%%56
\bibitem{guh} G\" uhne, O. Characterizing entanglement via uncertainty relations. \emph{Phys. Rev. Lett.} \textbf{2004}, \emph{9}, 117903.
%%57
\bibitem{giov} Giovannetti, V.; Lloyd,   S.; Maccone, L. Advances in quantum metrology. \emph{Nat. Photonics} \textbf{2011}, \emph{5}, 222.
%%58
\bibitem{deba} Mondal, D.; Datta,  C.; Sazim, S.K. Quantum coherence sets the quantum speed limit for mixed states. \emph{Phys. Lett. A} \textbf{2016}, \emph{380}, 689--695.

\bibitem{marv} Marvian, I.; Spekkens, R.W.; Zanardi, P. Quantum speed limits, coherence, and asymmetry. \emph{Phys. Rev. A} 2016, 93, 052331.

\bibitem{deff} Deffner, S.; Campbell, S. Quantum speed limits: from Heisenberg's uncertainty principle to optimal quantum control. \emph{J. Phys. A} \textbf{2017}, \emph{50}, 453001.
%%61
\bibitem{pires}  Pires, D.P.; Cianciaruso, M.; Céleri, L.C.; Adesso, G.; Soares-Pinto, D.O. Generalized geometric quantum speed limits. \emph{Phys. Rev. X} \textbf{2016}, \emph{6}, 021031,

\bibitem{xiao}  Xiao, L. Experimental test of uncertainty relations for general unitary operators. \emph{Opt. Express} \textbf{2017}, \emph{25}, 17904--17910.

\bibitem{maw} Ma, W. Experimental test of Heisenberg's measurement uncertainty relation based on statistical distances. \emph{Phys. Rev. Lett.} \textbf{2016}, \emph{116}, 160405.
%%64
\bibitem{baek} Baek, S.Y.; ; Kaneda, F.; Ozawa, ; M.; Edamatsu, K. Experimental violation and reformulation of the Heisenberg's error disturbance uncertainty relation. \emph{Sci. Rep.} \textbf{2013}, \emph{3}, 2221.
%%67
\bibitem{mondal} Mondal, D.;  Bagchi,  S.; Pati,  A.K. Tighter uncertainty and reverse uncertainty relations. \emph{Phys. Rev. A} \textbf{2017}, \emph{95}, 052117.







\bibitem{schiff} Schiff, L. \textit{ Quantum Mechanics, International series in pure and applied physics}; McGraw-Hill: New York, NY, USA, 
 1955.

\bibitem{dj}  Griffiths, D. J. \textit{Introduction to Quantum Mechanics}; Pearson Prentice Hall: Upper Saddle River, NJ, USA, 
 2005.
%%59%
\bibitem{reif1} Reif, F. \textit{ Fundamentals of Statistical and Thermal Physics}; Waveland Press: Long Grove, IL, USA, 
 2009.
\bibitem{macc} Maccone, L.; Pati, A.K. Stronger uncertainty relations for all incompatible observables. \emph{Phys. Rev. Lett.} \textbf{2014}, \emph{113}, 260401.

\bibitem{pc}  Chattopadhyay, P.; Mitra,  A.; Paul,  G. Probing Uncertainty Relations in Non-Commutative Space. \emph{Int. J. Theor. Phys.} \textbf{2019}, \emph{58}, 2619--2631.


\bibitem{dn1}Cerone, P.; Dragomir, S.S. \textit{Mathematical Inequalities}; Chapman and Hall/CRC: London, UK,
 2011; pp. 241--313.


\bibitem{saham} Saha, M.;  and Srivastava, B. N. \textit{Treatise on Heat}; Longman, Rees, Orme, Brown, Green \& Longman: New York, NY, USA, 1935.

%\bibitem{reif}Reif, Frederick. ``Fundamentals of statistical and thermal physics". Waveland Press, 2009.

\bibitem{say} Saygin, H.; \c{S}i\c{s}man,  A.  Quantum degeneracy effect on the work output from a Stirling cycle. \emph{J. Appl. Phys.} \textbf{2001}, \emph{90}, 3086--3089.

\bibitem{agar} Agarwal, G.S.; Chaturvedi, S. Quantum dynamical framework for Brownian heat engines. \emph{Phys. Rev. E} \textbf{2013}, \emph{88}, 012130.

\bibitem{hua}  Huang, X.L.;  Niu, X.Y.;  Xiu, X.M.; Yi, X.X. Quantum Stirling heat engine and refrigerator with single and coupled spin systems. \emph{Eur. Phys. J. D} \textbf{2014}, \emph{68}, 32.

\bibitem{blick}  Blickle, V.; Bechinger, C. Realization of a micrometre-sized stochastic heat engine.  \emph{Nat. Phys.} \textbf{2011}, \emph{8}, 143.


\bibitem{thomas} Thomas, G.; Das,  D.; Ghosh,  S. Quantum heat engine based on level degeneracy. \emph{Phys. Rev. E} \textbf{2019}, \emph{100}, 012123.








\bibitem{nied} Niedenzu, W.;  Mukherjee, V.;  Ghosh, A.; Kofman, A.G.;  Kurizki, G. Quantum engine efficiency bound beyond the second law of thermodynamics.  \emph{Nat. Commun.} \textbf{2018}, \emph{9}, 165.

\bibitem{plen} Plenio, M.B.; Virmani, S. An introduction to entanglement measures. In \emph{Quantum Information and Coherence}; Springer: Cham, Switzerland, 2014; pp. 173--209

\bibitem{albe}Alberto, P.;  Fiolhais, C.; Gil, V.M.S. Relativistic particle in a box. \emph{Eur. J. Phys.} \textbf{1996}, \emph{17}, 19.

\bibitem{ryu} Ryu, S.; Takayanagi,  T. Holographic derivation of entanglement entropy from the anti–de sitter space/conformal field theory correspondence. \emph{Phys. Rev. Lett.} \textbf{2006}, \emph{96}, 181602.

\bibitem{kempf} Kempf, A.; Mangano,  G.; Mann, R.B. Hilbert space representation of the minimal length uncertainty relation. \emph{Phys. Rev. D} \textbf{1995}, \emph{52}, 1108.

\bibitem{ques} Quesne, C.; Tkachuk, V.M. Composite system in deformed space with minimal length. \emph{Phys. Rev. A} \textbf{2010}, \emph{81}, 012106.

\bibitem{melj} Meljanac, S.;  Kre{\v{s}}i{\'c}-Juri{\'c}, S. Noncommutative differential forms on the kappa-deformed space. \emph{J. Phys. A} \textbf{2009}, \emph{42}, 365204.


\bibitem{rezek}Rezek, Y.; Kosloff, R. Irreversible performance of a quantum harmonic heat engine. \emph{New J. Phys.} \textbf{2006}, \emph{8},  83.
\end{thebibliography}
\end{document}